\documentclass[pra,amsmath,amssymb,amsfonts,twocolumn,
    superscriptaddress,nobalancelastpage,floatfix]{revtex4}

\usepackage{bbm}
\usepackage{amssymb}
\usepackage{amsmath}
\usepackage{verbatim}
\usepackage{graphicx}
\usepackage{epsfig}
\usepackage{dsfont}
\usepackage{color}

    {\hspace*{\fill}$\Box$\vspace{1.5ex}\par}

\newcommand{\ad}{a^{\dagger}}
\newcommand{\al}{\alpha^{\dagger}}
\newcommand{\be}{\beta^{\dagger}}
\newcommand{\ga}{\gamma^{\dagger}}
\newcommand{\de}{\delta^{\dagger}}

\newcommand{\tr}{\mbox{tr}}

\newcommand{\ket}[1]{\left|#1\right\rangle}
\newcommand{\bra}[1]{\left\langle#1\right|}

\newcommand{\vn}{\vec n}

\begin{document}
\title{Fermionic Projected Entangled Pair States}
\author{Christina V.\ Kraus}
\affiliation{Max-Planck-Institute for Quantum Optics, Hans-Kopfermann-Str.\ 1, D-85748 Garching,
Germany.}
\author{Norbert Schuch}
\affiliation{Max-Planck-Institute for Quantum Optics, Hans-Kopfermann-Str.\ 1, D-85748 Garching,
Germany.}
\author{Frank Verstraete}
\affiliation{Fakult\"at f\"ur Physik, Universit\"at Wien, Boltzmanngasse 5, A-1090 Wien, Austria.}
\author{J.\ Ignacio Cirac}
\affiliation{Max-Planck-Institute for Quantum Optics, Hans-Kopfermann-Str.\ 1, D-85748 Garching,
Germany.}

\begin{abstract}
We introduce a family of states, the fPEPS, which describes fermionic systems on lattices in arbitrary
spatial dimensions. It constitutes the natural extension of another family of states, the PEPS, which
efficiently approximate ground and thermal states of spin systems with short-range interactions. We
give an explicit mapping between those families, which allows us to extend previous simulation methods
to fermionic systems. We also show that fPEPS naturally arise as exact ground states of certain
fermionic Hamiltonians. We give an example of such a Hamiltonian, exhibiting criticality while obeying
an area law.
\end{abstract}


\maketitle

\section{Introduction}Understanding the behavior of correlated quantum many-body systems is one of the
most challenging problems in various fields of physics. For spin systems on a lattice with local (i.e.,
short--range) interactions, powerful methods have been developed in recent years. They rely on families
of states which, on the one hand, depend on very few parameters and, on the other, approximate the
quantum state of the spins in thermal equilibrium. In one spatial dimension, Matrix Product States
(MPS)~\cite{mps-def} (which underly \cite{ostlund-rommer,frank:dmrg-mps} the successful Density Matrix
Renormalization Group algorithm~\cite{white:DMRG,schollwoeck:rmp}) provide a good approximation to the
ground state of any gapped local Hamiltonian. Projected Entangled Pair States
(PEPS)~\cite{frank:2D-dmrg,murg:peps-review} (cf.\ also~\cite{aklt,nishino:tps}), which naturally
extend MPS to higher spatial dimensions, approximate spin states at any finite
temperature~\cite{hastings:locally}, and have been successfully used to simulate spin systems which
cannot be dealt with
otherwise~\cite{isacsson:peps,murg:peps-alg,murg:J1J2,%
jordan:bosehubbard-ipeps}.

Fermionic quantum many-body systems are central to many of the most fascinating effects in condensed
matter physics. In one spatial dimension, it is possible to adapt the methods based on MPS to such
systems thanks to the Jordan-Wigner transform, which maps fermions into spins while keeping the
interactions local. In higher dimensions, however, this is no longer possible: fermionic operators at
different locations anticommute, which effectively induces nonlocal effects when mapping fermions to
spins. Thus, the use of PEPS to describe fermionic systems is no longer justified (see
however~\cite{frank:2d-jw,evenbly:fermion-mera} for different approaches).

In this article we introduce a new family of states, the fermionic Projected Entangled Pair States
(fPEPS), which naturally extend the PEPS to fermionic systems. According to their definition, fPEPS are
well suited to describe fermionic systems with local interactions. They can be, in turn, efficiently
described in terms of standard PEPS at the prize of having to double the number of parameters. This
automatically implies that the algorithms introduced to simulate ground and thermal states, as well as
the time evolution of spin systems using PEPS~\cite{frank:2D-dmrg,murg:peps-review}, can be readily
adapted to fPEPS.  We also show that certain fPEPS are exact ground states of local fermionic
Hamiltonians, in as much the same way as PEPS are for spins~\cite{perez-garcia:parent-ham-2d}.  In
particular, we give the explicit construction of a Gaussian Hamiltonian which has a fPEPS as exact
ground state. Remarkably, the state is critical, i.e. gapless with polynomially decaying correlations,
yet obeys an entropic area law \cite{cramer:arealaw-review}, in contrast to what happens with other
free fermion systems \cite{fermion-arealaw}.

We have organized this paper as follows. First, we will briefly review PEPS and explain why they are
well suited to describe spin systems with local interactions in thermal equilibrium. Then, we will
construct the family of fPEPS following the same idea. We will then consider a subfamily of fPEPS for
which we can build local ``parent'' Hamiltonians, i.e. those for which they are exact ground states.
Finally, we will give a particular example which presents criticality. For the sake of simplicity, we
will concentrate on two spatial dimensions.

\section{Construction of PEPS}
\begin{figure}[t]
\begin{center}
\includegraphics[width=0.9\columnwidth]{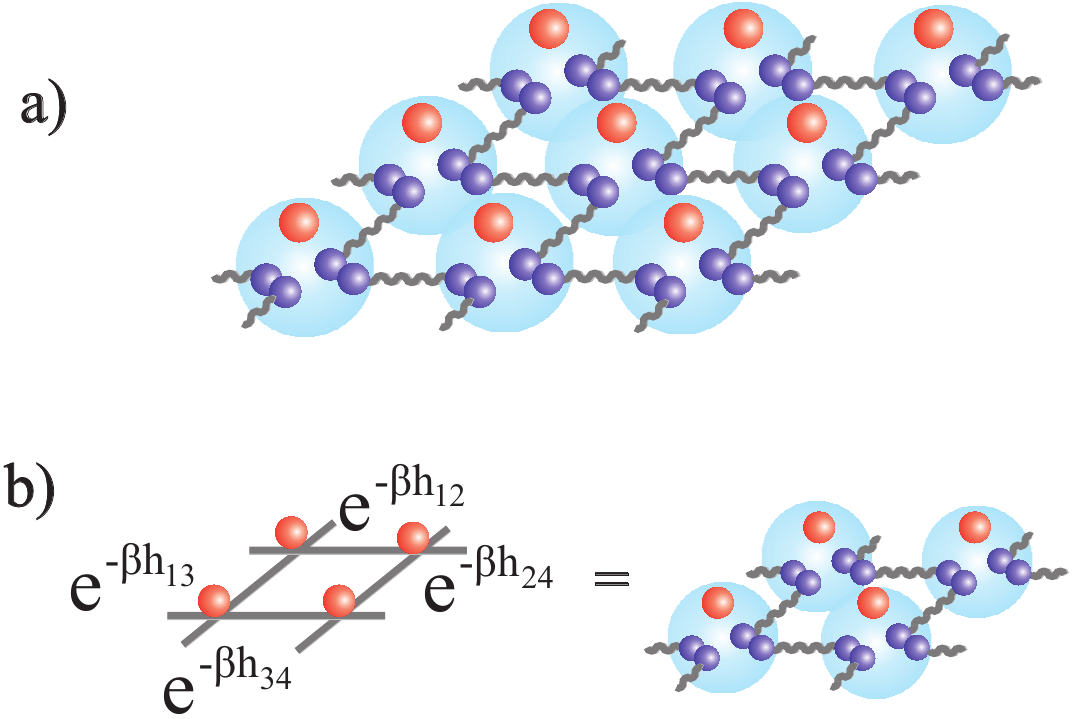}
\end{center}
\caption{\textbf{a)} Construction of a PEPS in two dimensions.  The balls joined by lines represent
pairs of maximally entangled $D$-dimensional auxiliary spins, which are then mapped to the physical
spins (red), as illustrated by the light blue spheres.  \textbf{b)} Why PEPS approximate thermal states
well: $\exp[-\beta h_{ij}]$ can be implemented using local maps only if an entangled pair is available.
\label{fig:peps} }
\end{figure}

For simplicity, let us consider a $2D$ lattice of $N \equiv N_h\cdot N_v$ spin 1/2 particles, with
states $|0\rangle$ and $|1\rangle$. To each node of coordinates $(h,v)$ we associate four auxiliary
spins, with states $|n\rangle$ ($n=0,\ldots,D-1$), where $D$ is called bond dimension. Each of them is
in a maximally entangled state $\sum|n,n\rangle$ with one of its neighbors, as indicated in
Fig.~\ref{fig:peps}a. The PEPS $|\Psi\rangle$ is obtained by applying a linear operator (``projector'')
to each node that maps the auxiliary spins onto the original ones. This operator can be parametrized as
\begin{equation}\label{Pmap}
P_{(h,v)}=
    \sum_{l,r,u,d=0}^{D-1}
    \sum_{k=0}^1(B_{(h,v)})^{[k]}_{l,r,u,d}
    |k\rangle\langle l,r,u,d|.
\end{equation}

Let us now explain why PEPS are well suited to describe spins in thermal equilibrium in the case of
local Hamiltonians, $H=\sum h_\lambda$. For simplicity, we will assume that each $h_\lambda$ acts on
two neighboring spins.  We first rewrite the (unnormalized) density operator $e^{-\beta
H}=\tr_B[\ket\chi\bra\chi]$, where $\ket\Psi=e^{-\beta H/2}\otimes\openone\ket\chi_{AB}$ is a
purification~\cite{verstraete:finite-t-mps} and $\ket{\chi}_{AB}$ a pairwise maximally entangled state
of each spin with another one, the latter playing the role of an environment. We will show now that
$\ket\Psi$ can be expressed as a PEPS.  We consider first the simplest case where
$[h_\lambda,h_{\lambda'}]=0$, so that $\ket\Psi = \prod_{\lambda}e^{-\beta
h_{\lambda}/2}\otimes\openone\ket\chi_{AB}$. The action of each of the terms $e^{-\beta h_{\lambda}/2}$
on two spins in neighboring nodes can be viewed as follows: we first include two auxiliary spins, one
in each node, in a maximally entangled state, and then we apply a local map in each of the nodes which
involves the real spin and the auxiliary spin, which ends up in $|0\rangle$. By proceeding in the same
way for each term $e^{-\beta h_{\lambda}/2}$, we end up with the PEPS description
(see~Fig.~\ref{fig:peps}b). This is valid for all values of $\beta$, in particular for
$\beta\to\infty$, i.e., for the ground state. In case the local Hamiltonians do not commute, a more
sophisticated proof is required~\cite{hastings:locally}. One can, however, understand qualitatively why
the construction remains to be valid by using a Trotter decomposition to approximate $ e^{-\beta H}
\approx \prod_{m=1}^M\prod_\lambda e^{-\beta h_{\lambda}/2M}$ with $M\ll 1$. Again, this allows for a
direct implementation of each $\exp[-\beta h_\lambda/2M]$ using one entangled bond, yielding $M$ bonds
for each vertex of the lattice.  Since, however, the entanglement induced by each $\exp[-\beta
h_\lambda/2M]$ is very small, each of these bonds will only need to be weakly entangled, and the $M$
bonds can thus be well approximated by a maximally entangled state of low dimension. Note that the
spins belonging to the purification do not play any special role in this construction, and thus we will
omit them in the following.

\section{Construction of fPEPS}%
We will now extend the above construction to fermionic systems, in such a way that the same arguments
apply. We consider fermions on a lattice, and work in second quantization. For a Hamiltonian $H=\sum
h_\lambda$, each term  $h_\lambda$ must contain an even number of fermionic operators, in order for the
Trotter decomposition to be still possible. Thus, we just have to find out how to express the action of
$e^{-\beta h_\lambda}$ in terms of auxiliary systems. This is very simple: one just has to consider
that the auxiliary particles are fermions themselves, forming maximally entangled states, and write a
general operator which performs the mapping as before. Following this route, we arrive at the
definition of fPEPS. More specifically, we define at each node $(h,v)$ four auxiliary fermionic modes,
with creation operators $\al_{(h,v)}, \be_{(h,v)},\ga_{(h,v)},\de_{(h,v)}$, respectively. We define
\begin{eqnarray}
\label{phih}
    H_{(h,v)} &=&
    \tfrac{1}{\sqrt{2}}
    (1 + \be_{(h,v)}\al_{(h+1,v)})\\
    V_{(h,v)} &=&
    \tfrac{1}{\sqrt{2}}
    (1 +\de_{(h,v)}\ga_{(h,v+1)})\label{phiv}
\end{eqnarray}
which create maximally entangled states out of the vacuum. We also define the ``projectors''
\begin{equation}\label{Px}
Q_{(h,v)}= \sum(A_{(h,v)})_{lrud}^{[k]} a^{\dagger k}_{(h,v)}
    \alpha_{(h,v)}^{l}
    \beta_{(h,v)}^{r}
    \gamma_{(h,v)}^{u}
    \delta_{(h,v)}^{d},
\end{equation}
where $a_{(h,v)}$ is the annihilation operator of the physical fermionic mode, and the sum runs for all
the indices from $0$ to $1$, with the condition that $(u+d+l+r+k)\,\mathrm{mod}\,2=c$, were $c$ is
fixed for
each node~\footnote{%
In fact one can freely choose $c$ for all but one $Q_{(h,v)}$: Since, e.g., the bond (\ref{phih}) is
invariant under $(i\beta_{(h,v)}+\beta_{(h,v)}^\dagger)
    (i\alpha_{(h+1,v)}+\alpha_{(h+1,v)}^\dagger)$,
the corresponding maps (\ref{Px}) can be right multiplied with it, switching their parity.}. The latter
is related to the parity of the $h_{\lambda}$ and will ensure that the parity of the fPEPS is well
defined. The fPEPS is then
\begin{equation}\label{DefPEPS}
|\Psi\rangle =\langle \prod_{(h,v)}Q_{(h,v)}
    \prod_{(h,v)} H_{(v,h)}
    V_{(v,h)}\rangle_{\rm aux} \; |{\rm vac}\rangle,
\end{equation}
where the expectation value is taken in the vacuum of the auxiliary modes, and $|{\rm vac}\rangle$
denotes the vacuum of the physical fermions. Note that the definition of fPEPS straightforwardly
extends to systems with both more than one physical mode per site and more than one mode per bond, as
well as to open boundaries or higher spatial dimensions.

\section{Relation between fPEPS and PEPS}%
\begin{figure}
\begin{center}
\includegraphics[width=0.9\columnwidth]{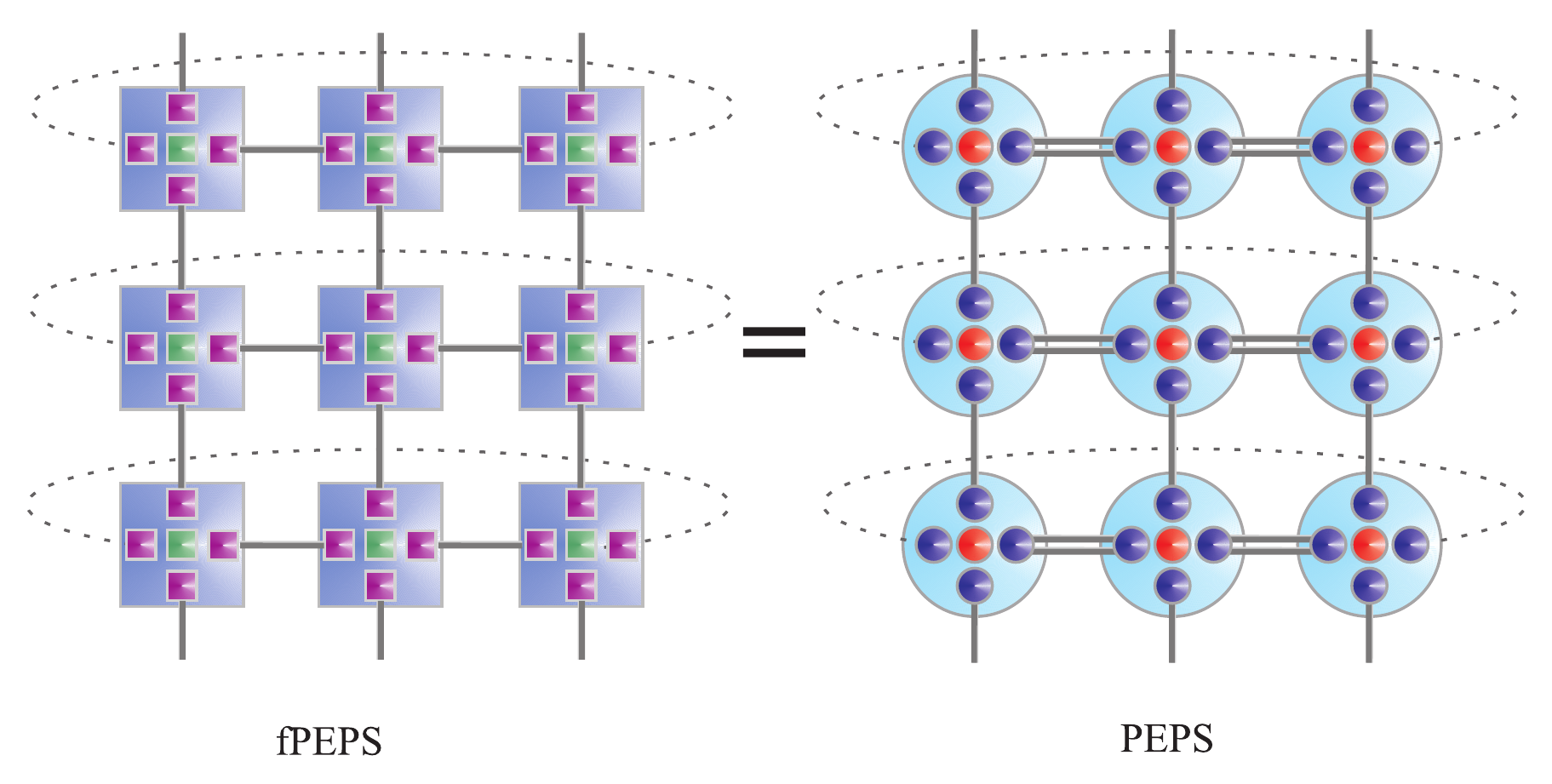}
\end{center}
\caption{ \label{fig:fpeps-peps} Every fPEPS can be represented as a PEPS at an extra cost of at most
one additional bond per link (shown for a $3\times 3$ PBC lattice).}
\end{figure}

Next, we will find an efficient description of any fPEPS in terms of standard PEPS. With that, one can
readily use the methods introduced for PEPS \cite{frank:2D-dmrg,murg:peps-review} in order to determine
physical observables, as well as to perform simulations of ground or thermal states, and time
evolution. We have to identify the Fock space of the fermionic modes with the Hilbert space of spins.
For that, we sort the lattice sites according to $M=(v-1)N_h + h$ and associate $a_1^{\dagger k_1}
\ldots a_N^{\dagger k_N}|{\rm vac}\rangle$ to the spin state $|k_1,\ldots,k_N\rangle$. Then we write
$|\Psi\rangle$ in that basis, and express it as a PEPS in terms of tensors $B$~(\ref{Pmap}). The goal
is to find the relation between the tensors $B$ (corresponding to the spin description) and $A$
(fermionic description). In principle, the fPEPS to PEPS transformation can be done straightforwardly
by adding extra bonds to the PEPS which take care of the signs which arise from reordering the
fermionic operators; however, this would lead to a linear number of bonds per link and thus to a
dimension which is exponential in $N$. Remarkably, it is possible to express every fPEPS as a PEPS by
introducing only \emph{one} additional bond per horizontal link as follows: Replace each fermionic bond
by a bond of maximally entangled spins, adding one additional horizontal qubit bond everywhere except
at the boundaries (see Fig.~\ref{fig:fpeps-peps}). This means that the tensor $B$ will have now two
more indices, say $l'$ and $r'$, which are associated to those new bonds. Then, we find the relation
\begin{equation}
\label{B1} (B_{h,v})^{[k]}_{lrr'ud}= (-1)^{f_{(h,v)}(k,u,d,l,r)}(A_{h,v})^{[k]}_{lrud}(-1)^{(d+l)r'}
\end{equation}
for $h=1$, while for $h>1$ we have
\begin{align}
&(B_{h,v})^{[k]}_{ll'rr'ud}= \label{B2}\\
&\quad(-1)^{f_{(h,v)}(k,u,d,l,r)} (A_{h,v})^{[k]}_{lrud}(-1)^{dr'}\delta_{l',(r'+u+d)
\,\mathrm{mod}\,2}\nonumber\ ,
\end{align}
where $f_{(h,v)}(k,u,d,l,r)$ is a function which only depends on the local indices, and $r'=0$ for
$h=N_h$.

Let us briefly explain how to obtain this result. Consider an fPEPS of the form \eqref{DefPEPS} which
we want to bring into the normal ordered form by commuting the fermionic operators.  To this end, we
perform the following three steps on the total projector $\prod Q_{(h,v)}$, observing that local sign
contributions can be absorbed in the tensors $(A_{(h,v)})^{[k]}_{lrud}$: First, commute all physical
modes to the left. This results in a factor $(-1)^{p(p-1)/2}$, where $p=\sum_{(h,v)}k_{(h,v)}$ is the
parity of the fPEPS; since the latter is fixed, this yields a global phase. Next, contract the
horizontal bonds: The non-boundary bonds only yield local contributions, while the horizontal boundary
bond on any line $v$ gives a contribution $(-1)^{l_{(1,v)}\Pi(1,v)}$ with
$\Pi(h,v)=\sum_{j>h}(u_{(j,v)}+d_{(j,v)})$. Finally, contract the vertical bonds, proceeding
column\-wise from $h=1$. For each bond between $(h,v)$ and $(h,v+1)$ this gives a sign contribution
$(-1)^{d_{(h,v)}\Pi(h,v)}$; due to the fixed parity of the bonds this holds even for the bonds across
the boundary. Thus, all signs can be computed if the respective parity $\Pi(h,v)$ is available at each
site, which is achieved by the additional bonds passing this information to the left. Note that the
same proof applies to open boundaries, as well as systems with more physical or virtual modes per site,
without the need for further extra bonds to compute $\Pi(h,v)$. Similarly, one can derive a
corresponding result for higher dimensions.

\section{Fermionic Gaussian states and parent Hamil\-to\-ni\-ans}%
Fermionic Gaussian states~\cite{bravyi:flo} (also known as quasi-free states) constitute an important
subclass of states, as they appear as ground and thermal states of quadratic Hamiltonians,
corresponding to free fermion or BCS states. These states can be written as an exponential of a
quadratic form in the fermionic operators, and are thus completely characterized by their covariance
matrix $\Gamma_{kl}^{(x,y)}=\tr[\tfrac i2 [c_k^{(x)},c_l^{(y)}]\rho]$, where $c_{i}^{(1)}=\ad_i + a_i$
and $c_{i}^{(2)}=(-i)(\ad_i - a_i)$ are Majorana operators. We will now introduce Gaussian fPEPS, which
we then use to show that fPEPS naturally appear as ground states of free local Hamiltonians. The
techniques used here follow closely the corresponding methods for bosons introduced
in~\cite{schuch:GMPS}.

Gaussian fPEPS are obtained by restricting the map (\ref{Px}) to be Gaussian ($H$ and $V$ are already
of that form). Those transform Gaussian states into Gaussian states, so that they can be characterized
through the map $\Gamma_\mathrm{in}\to\Gamma_\mathrm{out}$. The most general (pure) map can be written
as \cite{bravyi:flo} $\Gamma_\mathrm{out}= B(D-\Gamma_\mathrm{in})^{-1}B^T + A$ with
\begin{equation}
G = \left(%
\begin{array}{cc}
  A & B \\
  -B^T & D \\
\end{array}%
\right)=-G^{T}\,, \;\;GG^T = -\openone\ .
\end{equation}
We denote the CM of the translationally invariant states of the virtual modes by
$\Gamma_\mathrm{in}=\oplus\omega_{h,v}$ where $\omega_{h,v}$ is the CM of the maximally entangled
horizontal resp. vertical bonds. Then the desired family of states can be obtained by applying the same
Gaussian map to each node $\vec{n}= (h,v)$ of the lattice: $\mathcal G=\oplus_{\vec{n}}
\tilde{\mathcal{G} }$, where $\tilde{\mathcal{G}}\tilde{\mathcal{G}}^T = \mathds{1}$.
Due to translational invariance, $\Gamma_\mathrm{out}$ can be conveniently expressed in Fourier space,
$\Gamma_\mathrm{out}=\oplus_{\vec\phi}\hat\Gamma_\mathrm{out}(\vec\phi)$, with $
\hat\Gamma_\mathrm{out}= B [D-\hat\omega(\vec\phi) ]^{-1}B^T + A$, where $\vec\phi=(\tfrac{2\pi
k_h}{N_h},\tfrac{2\pi k_v}{N_v})$ is the reciprocal lattice vector.  As we show in Appendix \ref{sec:Gaussian}, it is straightforward to see that
the $\vec\phi$-dependence of $\hat\omega(\vec\phi)$ yields
\begin{align}\label{eq:ft-state}
&\hat{\Gamma}_{\mathrm{out}}(\vec{\phi})= \\
&\frac{1}{d(\vec{\phi})} \left(
                             \begin{array}{cccc}
                               0 & \mathrm{Re}(q(\vec{\phi})) & -\mathrm{Im}(q(\vec{\phi})) & p(\vec{\phi}) \\
                               -\mathrm{Re}(q(\vec{\phi})) & 0 & p(\vec{\phi}) & \mathrm{Im}(q(\vec{\phi}))\\
                               \mathrm{Im}(q(\vec{\phi})) & -p(\vec{\phi}) & 0 & \mathrm{Re}(q(\vec{\phi})) \\
                               -p(\vec{\phi}) & -\mathrm{Im}(q(\vec{\phi})) & -\mathrm{Re}(q(\vec{\phi})) & 0 \\
                             \end{array}
                           \right).\nonumber
\end{align}

with $p$, $q$ and $d$ low-degree polynomials in $\vec\phi$; in particular, $d(\vec\phi) =
\det[D-\hat{\omega}(\vec\phi)]$. Now define the Hamiltonian $H = i\sum_{kl}h_{kl}c_kc_l$, where $h$ is
defined through its Fourier transform $\hat h(\vec\phi)=d(\vec\phi)\hat\Gamma_\mathrm{out}(\vec\phi)$.
$H$ has $\Gamma_\mathrm{out}$ as its ground state, since $\hat\Gamma_\mathrm{out}(\vec\phi)$ and $\hat
h(\vec\phi)$ are diagonal in the same basis, and unless $H$ is gapless---corresponding to zeros of
$d(\vec\phi)$---the ground state is unique. Moreover, since the degree of $p$ and $q$ is bounded by
twice the number of virtual modes per site, it follows that $H$ is local. 

\begin{figure}[t]
\parbox{3.8cm}{
\parbox{3.8cm}{\mbox{\textbf{a)}\hspace*{3.2cm}}\textbf{b)}}\\
\includegraphics[width=3.6cm]{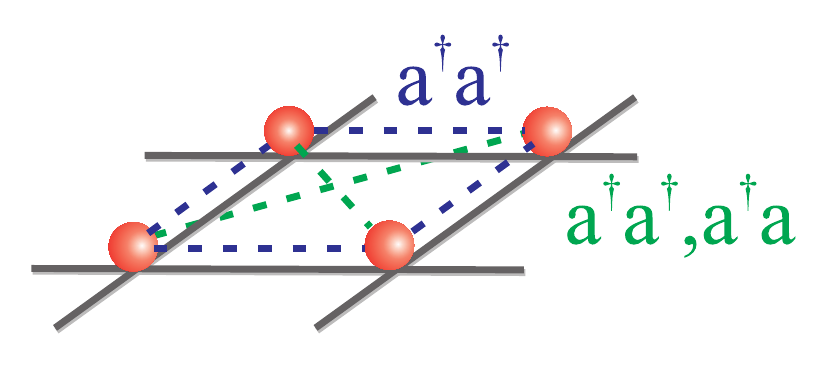}
\rule{0cm}{.8cm}\\
\parbox{3cm}{\textbf{c)\hspace*{\fill}}}\\
\includegraphics[width=2.6cm]{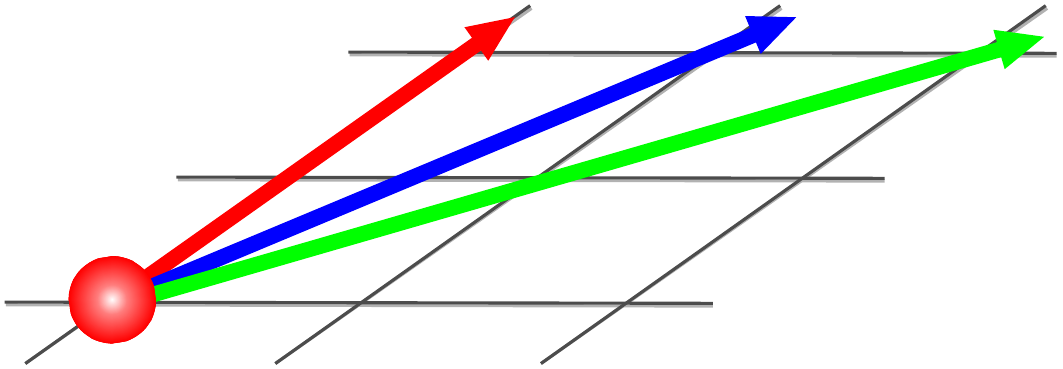}\\
\rule{0cm}{.8cm} }
\parbox{4.5cm}{
\includegraphics[width=4.5cm]{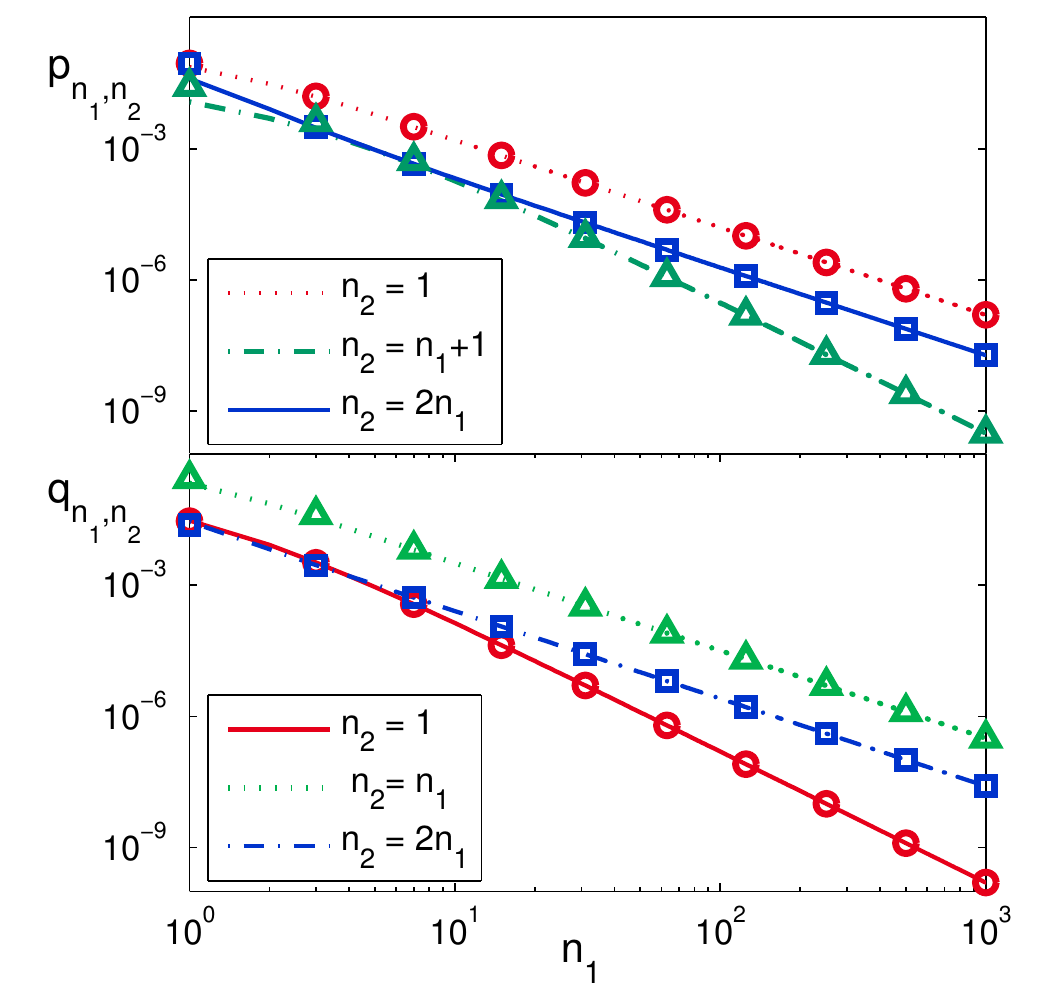}
}
\caption{ \label{fig3} \textbf{a)} Hopping terms in the Hamiltonian. \textbf{b)} Exact value and
asymptotic scaling for the correlations in direction of the axis (red), along the diagonal (green) and
along the direction $(n,2n)$ (blue) [cf.\ \textbf{c)}], for Majorana operators of the same type (top)
and different type (bottom).}
\end{figure}

\textbf{Example} Let us now give an example of a local Hamiltonian which has an fPEPS as its exact ground state. We present only the main results in the main text, and refer the reader interested in the details to Appendix \ref{sec:example}. We
choose the (translational invariant) fPEPS projector $Q=e^{
    (i\alpha+\beta)(-\gamma+i\delta)+\alpha\beta+\gamma\delta+
    a^\dagger(-i\alpha-\beta-\gamma+i\delta)
    }$
which yields $p(\vec\phi)/d(\vec\phi)=(\sin\phi_1-\sin\phi_2)/(-1+\sin\phi_1\sin\phi_2)$ and
$q(\vec\phi)/d(\vec\phi)= \cos\phi_1\cos\phi_2/(-1+\sin\phi_1\sin\phi_2)$. The resulting parent
Hamiltonian is (Fig.~\ref{fig3}a)
\begin{eqnarray*}\label{Hcrit}
H_\mathrm{crit} &=&
   2i \sum_{(h,v)}
    \ad_{(h,v)}\ad_{(h,v+1)}
    -\ad_{(h,v)}\ad_{(h+1,v)}
    +\mathrm{h.c.}
\\
    &&-\sum_{(h,v)}
    \ad_{(h,v)} (a_{(h+1,v+1)}+a_{(h+1,v-1)})+\mathrm{h.c.}
\end{eqnarray*}
and $N_h,N_v$ odd, which will ensure that the ground state is unique. By Fourier transforming
$\hat\Gamma_\mathrm{out}(\vec\phi)$ into position space, one obtains that the correlations of Majorana
operators of equal (different) type at distance $(n_1,n_2)$ scale asymptotically as the real
(imaginary) part of $K(n_1,n_2)=(n_1+3+in_2)/{(n_1+1+in_2)^3}$ for $n_1+n_2$ odd (even) and vanish
otherwise (Fig.~\ref{fig3}b). Notably, the ground state possesses correlations that decay as power laws
and the Hamiltonian is gapless in the limit $N\to\infty$. In fact, our example provides us with a
critical fermionic system obeying the area law, which directly follows from the fact that its ground
state is a PEPS with bounded bond dimensions. Note that, although $H_{\rm {crit}}$ is not particle
conserving, it can be converted into a particle conserving one via a simple particle--hole
transformation in the B sublattice. This new Hamiltonian possesses a spectrum with a Dirac point
separating the modes with positive and negative energies. Thus, the Fermi surface has zero dimension,
which explains why our results do not contradict the violation of the area law expected for free
Fermionic systems \cite{fermion-arealaw}.

In summary, in this work we have introduced fermionic PEPS (fPEPS) which are obtained by applying
fermionic linear maps to maximally entangled fermionic states placed between nearest neighbors. This
construction resembles the construction of PEPS and is well suited to describe ground and thermal
states of local fermionic Hamiltonians (both free and interacting), in the same way as PEPS are suited
to describe ground states of local spin systems. We have then shown how fPEPS can be transformed into
PEPS at the cost of only one additional bond providing an explicit mapping for the corresponding
tensors. This also demonstrates the use of fPEPS for numerical simulations. Further, we have
investigated the role of fPEPS as ground states of local Hamiltonians. To this end, we have introduced
Gaussian fPEPS and shown that they naturally arise as ground states of quasi-free local Hamiltonians.
Finally, we have used these tools to demonstrate the existence of local free fermionic Hamiltonians
which are critial without violating the area law.\\
 \emph{Note added:} After submission of this manuscript several algorithms based on
fPEPS have been developed and applied to interacting fermions~\cite{vidal:fPEPS, Iztok, Zhou}.

\begin{acknowledgments}
This work has been supported by the EU projects QUEVADIS  and SCALA, the DFG
Forschergruppe 635, the excellence cluster Munich Advanced Photonics (MAP), SFB Foqus and the Elite
Network of Bavaria programme QCCC.
\end{acknowledgments}

\begin{appendix}
\section{Gaussian fPEPS and Parent Hamiltonians\label{sec:Gaussian}}

In this Section we present the details that lead to Eq. (9). Recall, that we apply the translationally
invariant channel, Eq.(8), to the translÇationally invariant input state
$\Gamma_\mathrm{in}=\oplus\omega_{h,v}$. The structure of the problem suggests an approach in Fourier
space, and we introduce the Fourier transform of the the mode operators $\hat{f}_{\vec{\phi}} =
\left(\frac{1}{\sqrt{N}}\right)^2 \sum_{\vn}e^{-\frac{2\pi}{N}i \vec{\phi} \cdot \vn}f_{\vn}$, where
$f$ is either a physical or virtual mode, and $\vec\phi=(\tfrac{2\pi k_h}{N_h},\tfrac{2\pi k_v}{N_v})$
is the reciprocal lattice vector. Now we consider the CM of the output state in the qp-ordered form,
i.e. we write
\begin{equation}
\Gamma_{\mathrm{out}} = \left(
                 \begin{array}{cc}
                   \Gamma^{(1,1)}_{\mathrm{out}} & \Gamma^{(1,2)}_{\mathrm{out}} \\
                  \Gamma^{(2,1)}_{\mathrm{out}} & \Gamma^{(2,2)}_{\mathrm{out}} \\
                 \end{array}
               \right),
\end{equation}
where $\Gamma^{(r,s)}_{\mathrm{out}} = \langle \frac{i}{2}[c^{(r)},c^{(s)}]\rangle$, $r, s = 0,1$. The
translationally invariant construction is reflected in the fact that the blocks
$\Gamma^{(r,s)}_{\mathrm{out}}$ are circulant matrices. Hence, they all can be diagonalized
simultaneously by a Fourier transformation $\mathcal{F}$. The Fourier transform of
$\Gamma_{\mathrm{out}}$, $\hat{G}_{\mathrm{out}} = \mathcal{F}
\Gamma_{\mathrm{out}}\mathcal{F}^{\dagger}$, has diagonal blocks
\begin{eqnarray*}
\hat{G}_{\mathrm{out}}^{(r,s)} &=& \mathcal{F} \langle \frac{i}{2}[c^{(r)},c^{(s)}]\rangle
\mathcal{F}^{\dagger} =
\langle \frac{i}{2}[\hat{d}^{(r)},\hat{d}^{(s)\dagger}]\rangle\\
&=& \mathrm{diag}\left( g^{(r,s)}(\vec{\phi})\right),
\end{eqnarray*}
where $g^{(r,s)}(\vec{\phi}) \in \mathds{C}$  are the eigenvalues of the blocks $
\Gamma^{(r,s)}_{\mathrm{out}}$. The operators $\hat{d}^{(r)}_{\vec{\phi}}$ are the Fourier transformed
Majorana operators, $\hat{d}_{\vec{\phi}}^{(r)} = \left(\frac{1}{\sqrt{N}}\right)^2
\sum_{\vn}e^{-\frac{2\pi}{N}i \vec{\phi} \cdot \vn}c_{\vn}^{(r)}$, while the Majorana operators in the
reciprocal lattice space are given by $\hat{c}_{\vec{\phi}}^{(1)} = \hat{a}^{\dagger}_{\vec{\phi}} +
\hat{a}_{\vec{\phi}}$, $\hat{c}_{\vec{\phi}}^{(2)} = (-i)(\hat{a}^{\dagger}_{\vec{\phi}} -
\hat{a}_{\vec{\phi}}$), with CM $(\hat{\Gamma}_{\mathrm{out}}^{(x,y)})_{\vec{\phi}_1, \vec{\phi}_2} =
\langle\frac{i}{2}[\hat{c}_{\vec{\phi}_1}^{(x)}, \hat{c}_{\vec{\phi}_2}^{(y)}] \rangle$. Both
representations are linked via a unitary transformation. In the following we make use of
$\hat{G}_{\mathrm{out}}$ to derive properties of $\hat{\Gamma}_{\mathrm{out}}$. To this end, we regroup
the modes such that $\hat{G}_{\mathrm{out}} = \bigoplus_{\vec{\phi}}\hat{G}_{\mathrm{out}}(\vec{\phi})$
is a direct sum of blocks corresponding to the same lattice vector, i.e. we write
\begin{equation}
\hat{G}_{\mathrm{out}}(\vec{\phi}) = \left(
                                   \begin{array}{cc}
                                     g^{(1,1)}(\vec{\phi}) & g^{(1,2)}(\vec{\phi}) \\
                                     g^{(2,1)}(\vec{\phi}) & g^{(2,2)}(\vec{\phi}) \\
                                   \end{array}
                                 \right).
\end{equation}
Since $\Gamma_{\mathrm{out}}$ is antisymmetric and corresponds to a pure state, i.e.
$\Gamma_{\mathrm{out}}^2 = -\mathds{1}$, and the Fourier transformation is unitary, we find that $\hat
G_\mathrm{\mathrm{out}}(\vec\phi)$ can be written as

\begin{equation}
\label{eq:ft-state} \hat G_\mathrm{\mathrm{out}}(\vec\phi)
    = \frac{1}{d(\vec{\phi})}\left(%
\begin{array}{cc}
  i p(\vec\phi) & q(\vec\phi) \\
  -q(\vec\phi) & - i p(\vec\phi) \\
\end{array}%
\right),
\end{equation}
where $p(\vec{\phi})$, $q(\vec{\phi})$, $d(\vec{\phi}) \in \mathds{R}$. To obtain more information on
theses functions, we use the fact that the channel $\mathcal E$ describes a translationally invariant
map. This implies that the blocks $A$, $B$ and $D$ are block diagonal, and thus commute with the
Fourier transform. Hence,
\begin{equation}
\hat{G}_{\mathrm{out}} = \mathcal{F}\Gamma_{\mathrm{out}}\mathcal{F}^{\dagger} = B(D -
\hat{G}_{\mathrm{in}}(\vec{\phi}))^{-1}B^T + A,\label{eq:FT_Gamma}
\end{equation}
where $\hat{G}_{\mathrm{in}} = \mathcal{F} G_{\mathrm{in}} \mathcal{F}^{\dagger}$. We use that $(D -
\hat{G}_{\mathrm {in}}(\vec{\phi}))^{-1} = \mathrm{adj}(D - \hat{G}_{\mathrm{in}}(\vec{\phi}))/\det(D -
\hat{G}_{\mathrm{in}}(\vec{\phi}))$ where $\mathrm{adj}$ denotes the adjugate matrix, and we define
$d(\vec{\phi}) = \det(D - \hat{G}_{\mathrm{in}}(\vec{\phi}))$. As $\Gamma_{\mathrm{in}}$ is the
covariance matrix of a system of maximally entangled states between nearest neighbors, its Fourier
transform $\hat{G}_{\mathrm {in}}(\vec{\phi})$ is built out of terms of the form $e^{i\phi_{1,2}}$
only. Thus, $d(\vec{\phi}) = \det(D - \hat{G}_{\mathrm{in}}(\vec{\phi}))$ and $\mathrm{adj}(D -
\hat{G}_{\mathrm{in}}(\vec{\phi}))$  are polynomials of low order in $\phi_{1,2}$. As $B$ and $A$ are
local operators, we see that $p$ and $q$ are polynomials of low degree as well. These results lead to
the $\hat{\Gamma}_{\mathrm{out}}$ given in Eq. (9).

\section{Example of a critical fPEPS}\label{sec:example}
Like every Gaussian map the projector $Q=e^{ (i\alpha+\beta)(-\gamma+i\delta)+\alpha\beta+\gamma\delta+
    a^\dagger(-i\alpha-\beta-\gamma+i\delta)
    }$ can be described as a channel of the form given in Eq. (8), where
\begin{eqnarray*}\label{channel}
B &=& \frac{1}{2}\left(
        \begin{array}{cccccccc}
          1 & -1 & -1 & 1 & 0 & 0 & 0 & 0 \\
          0 & 0 & 0 & 0 & 1 & -1 & -1 & -1 \\
        \end{array}
      \right),\nonumber\\\nonumber\\
D &=& \frac{1}{4}\left(
        \begin{array}{cccccccc}
          0 & 0 & 2 & 2 & 1 & -1 & 1 & -1 \\
          0 & 0 & 2 & 2 & -1 & 1 & -1 & 1 \\
          -2 & -2 & 0 & 0 & 1 & -1 & 1 & -1 \\
          -2 & -2 & 0 & 0 & -1 & 1 & -1 & 1 \\
          -1 & 1 & -1 & 1 & 0 & 0 & 2 & 2 \\
          1 & -1 & 1 & -1 & 0 & 0 & 2 & 2 \\
         -1 & 1 & -1 & 1 & -2 & -2 & 0 & 0 \\
          1 & -1 & 1 & -1 & -2 & -2 & 0 & 0 \\
        \end{array}
      \right),
\end{eqnarray*}
and $A=0$. Using this representation, a straightforward calculation shows that the functions $p$, $q$,
and $d$ defined in Eq. (9) are of the form
$p(\vec\phi)/d(\vec\phi)=(\sin\phi_1-\sin\phi_2)/(-1+\sin\phi_1\sin\phi_2)$ and
$q(\vec\phi)/d(\vec\phi)= \cos\phi_1\cos\phi_2/(-1+\sin\phi_1\sin\phi_2)$.

Note that the success probability of the PEPS projection is related to the
absolute value of~[22]
\[
\mathrm{det}(D-\hat{G}_\mathrm{in}(\vec\phi))
\propto(-1-\sin\phi_1\sin\phi_2)\sin^2\tfrac{\phi_1}{2}\sin^2\tfrac{\phi_2}{2}\ ,
\]
which means that the fPEPS has zero norm (i.e., is not properly defined)
for $\vec\phi=(\tfrac\pi2,\tfrac{3\pi}2)$ and
$\vec\phi=(\tfrac{3\pi}{2},\tfrac{\pi}{2})$, as well as for $\phi_1=0,\pi$
and $\phi_2=0,\pi$. The former condition implies that the state is not
defined if the lattice size is a multiple of four in both directions. This
condition cannot be removed, since it is inherent to the way the critical
model is constructed -- these are exactly the zeros of $d(\vec\phi).$ The
other zeros, however, cancel out in (\ref{eq:FT_Gamma}), and it turns out
that one can modify the fPEPS construction to have nonzero norm in those
cases, without changing the CM of the state itself. This can be seen by
expressing the virtual fermions in terms of two Majorana modes: one finds
that only one of these modes per virtual fermion is connected to the
physical fermion by the PEPS projector, while the other is only perfectly
correlated with the corresponding Majorana mode of the opposite fermion.
Thus, these ``unused'' Majorana modes from perfectly correlated loops
around the torus, which make the state vanish for even loop sizes due to
the fermionic statistics. By properly modifying the bond across the
boundary in the unused Majorana mode one can prevent the state from
vanishing without affecting the fPEPS itself, which is still described by
(\ref{eq:FT_Gamma}).

Let us now show that the system is critical by deriving
the asymptotic behavior of the correlation functions $p$ and $q$ in position
space. For large systems we can replace the discrete Fourier transform by a continuous one. Let $\xi =
p,q$ and define
\begin{multline*}
\xi_{n_1,n_2} \equiv  \langle i
c^{(1)}_{(1,1)}c^{(y_{\xi})}_{(n_1, n_2)}\rangle =\\
\frac{1}{(2\pi)^2}\int_{0}^{2\pi}
\frac{\xi(\phi_1,\phi_2)}{d(\phi_1,\phi_2)}e^{in_1\phi_1}e^{in_2\phi_2}d\phi_1 d\phi_2,
\end{multline*}
where $y_{p} = 1$ and $y_{q} = 2$ for $p$- and $q$-correlations respectively. We make the substitution
$z=e^{i\phi_1}$ so that $dz = iz d\phi_1$, and arrive at
\begin{eqnarray}
\frac{p(z,\phi_2)}{d(z,\phi_2)} &=& i\frac{z^2-1-2iz\sin \phi_2}{2iz-(z^2-1)\sin\phi_2},\label{eq:p/d}\\
\frac{q(z,\phi_2)}{d(z,\phi_2)} &=& i\frac{(z^2+1)\cos \phi_2}{2iz-(z^2-1)\sin\phi_2}.
\end{eqnarray}
Then $\xi_{n_1,n_2}$ can be written as
\begin{eqnarray}\label{keq}
\xi_{n_1,n_2} &=& \frac{1}{2\pi}\int_0^{2\pi}d\phi_2 e^{in_2\phi_2}I_{n_1,n_2}^{(\xi)}(\phi_2),\\
I_{n_1,n_2}^{(´\xi)}(\phi_2) &=& \frac{1}{2\pi i}\oint_C dz
\frac{\xi(z,\phi_2)}{d(z,\phi_2)}z^{n_1-1},\nonumber
\end{eqnarray}
where $C$ is the closed loop on the unit circle in the complex plane. Since $d(z,\phi_2)^{-1}$ has
poles at $z_{\pm} = i\left(1\pm |\cos\phi_2|\right)/\sin\phi_2$ and $p$ as well as $q$ are holomorphic
at $z_{\pm}$, the integral $I_{n_1,n_2}^{(\xi)}(\phi_2)$ is proportional to the residue within $C$
according to the residue theorem. As only $z_-$ lies within the unit circle we obtain
\begin{equation}
I_{n_1,n_2}^{(\xi)}(\phi_2) = \frac{\xi(z_0,\phi_2)}{\partial_z d(z,\phi_2)|_{z=z_0}}.
\end{equation}
Next, we calculate the $p$-correlations. With the help of \eqref{eq:p/d} we obtain
\begin{eqnarray*}
I_{n_1,n_2}^{(p)}(\phi_2) &=& \frac{1}{2\pi i}\oint_C dz \;i\frac{z^2-1-2iz\sin
\phi_2}{2iz-(z^2-1)\sin\phi_2} z^{n_1-1}\nonumber\\
&=& i^{n_1+1}(1-|\cos\phi_2|)^{n_1}\frac{|\cos\phi_2|}{(\sin\phi_2)^{n_1+1}}.
\end{eqnarray*}
From the symmetry $I_{n_1,n_2}^{(p)}(\phi_2+\pi) = (-1)^{n_1+1}I_{n_1,n_2}^{(p)}(\phi_2)$ and the fact
that for $\phi_2 \in [-\pi/2,\pi/2]$ we have $|\cos\phi_2|=\cos\phi_2$, implying
\begin{equation}
I_{n_1,n_2}^{(p)}(\phi_2) =
i^{n_1+1}\frac{\cos\phi_2}{1-\cos\phi_2}\left(\tan\frac{\phi_2}{2}\right)^{n_1+1},
\end{equation}
we can conclude that
\begin{multline*}
p_{n_1,n_2} =\\ \frac{1}{2\pi}(1-(-1)^{n_1+n_2})2\mathrm{Re}
\left[\int_0^{\pi/2}d\phi_2\;e^{in_2\phi_2}I_{n_1,n_2}^{(p)}(\phi_2)\right].
\end{multline*}
To obtain that result, we have further made use of the relation $I_{n_1,n_2}^{(p)}(-\phi_2) =
(-1)^{n_1+1}I_{n_1,n_2}^{(p)}(\phi_2)$, so that for $n_1$ even (odd) $I_{n_1,n_2}^{(p)}(\phi_2)$ is an
odd (even) function, and only the sine (cosine)-part of the exponential $e^{in_2\phi_2}$ gives a
non-vanishing contribution. Following a similar strategy, one can derive

\begin{multline*}
q_{n_1,n_2} =\\ -\frac{1}{2\pi}(1+(-1)^{n_1+n_2})2\mathrm{Re}
\left[\int_0^{\pi/2}d\phi_2\;e^{in_2\phi_2}I_{n_1,n_2}^{(p)}(\phi_2)\right].
\end{multline*}

To prove criticality we are interested in the asymptotic behavior of the integral
\begin{equation*}
J_{n_1,n_2}=\int_0^{\pi/2}d\phi_2\;e^{in_2\phi_2}I_{n_1,n_2}^{(p)}(\phi_2).
\end{equation*}
The correlations are symmetric under the exchange of $n_1$ and $n_2$. This follows from translational
invariance and can also be seen directly from the form of $p(\phi_1,\phi_2)$ and $q(\phi_1,\phi_2)$.
Hence, to determine the asymptotic behavior, we can assume wlog. $n_1 \gg 1$. In this limit, the
absolute value of $I_{n_1,n_2}^{(p)}(\phi_2)$ attains its maximum for $\phi_2 = \pm
\mathrm{arccos}(1/n-1) \rightarrow \pi/2 $. We rewrite
$$I_{n_1,n_2}^{(p)}(\phi_2) = i^{n_1+1}c(\phi_2)e^{(n_1+1)t(\phi_2)},$$
where the functions $c$ and $t$ are given by $c(\phi_2) = \frac{\cos\phi_2}{1-\cos\phi_2}$ and
$t(\phi_2) = \log \left(\tan \frac{\phi_2}{2}\right)$. Next, we expand $c(\phi_2)$ and $t(\phi_2)$
around $\pi/2$: $\frac{\cos \phi_2}{1-\cos \phi_2} = -\left(\phi_2 - \frac{\pi}{2}\right) +
\left(\phi_2 - \frac{\pi}{2}\right)^2 + \mathcal{O}\left((\phi_2 - \frac{\pi}{2})^3\right)$, $
\log\left( \tan \frac{\phi}{2}\right) = \left(\phi_2 - \frac{\pi}{2}\right) + \frac{1}{6}\left(\phi_2 -
\frac{\pi}{2}\right)^3 + \mathcal{O}\left((\phi_2 - \frac{\pi}{2})^5\right)$.
Substituting $\phi_2 \rightarrow \phi_2 - \frac{\pi}{2}$ the integral attains the form $
J_{n_1,n_2}=i^{n_1+n_2+1}\int_{-\pi/2}^0d\phi_2 J(n_1,n_2,\phi_2)$ with kernel
\begin{multline*}
J(n_1,n_2,\phi_2)\\ = e^{in_2\phi_2} \left(-\phi_2 + \phi_2^2 \right)e^{(n_1+1)\phi_2 +
\phi_2^3/6(n_1+1)}(1 + \mathcal{O}(\phi_2^3)).
\end{multline*}
We use $J_{n_1,n_2} = \int_{-\infty}^0d\phi_2  J(n_1,n_2,\phi_2) - \int_{-\infty}^{-\pi/2}d\phi_2
J(n_1,n_2,\phi_2)$, and obtain
\begin{eqnarray*}
\int_{-\infty}^0d\phi_2  J(n_1,n_2,\phi_2) = \frac{3+n_1+in_2}{(1+n_1+in_2)^3} +
\mathcal{O}\left(\frac{1}{n_1^4},\frac{1}{n_2^4}\right),
\end{eqnarray*}
while the second integral can be bounded by

\begin{multline*}
\left|\int_{-\infty}^{-\pi/2}d\phi_2  J(n_1,n_2,\phi_2)\right| \leq e^{-(n_1+1)\pi/2} \times\\
\left| \int_{-\infty}^{-\pi/2}d\phi_2 e^{1/6(n_1+1)\phi_2^3}\left(-\phi_2 + \phi_2^2 \right)\left(1 +
\mathcal{O}\left(\phi_2^3\right)\right)\right|.
\end{multline*}
This gives rise to only an exponentially small correction that can be neglected in the asymptotic
limit. Summarizing, we see that the $p$-correlations are non-vanishing only for $n_1+n_2$ odd, while
$q$-correlations are non-vanishing only for $n_1+n_2$ even:
\begin{eqnarray*}
p_{n_1,n_2} &\sim& (1-(-1)^{n_1+n_2})\, \mathrm{Re}\left(\frac{3+n_1+in_2}{(1+n_1+in_2)^3}\right),\\
q_{n_1,n_2} &\sim& (1+(-1)^{n_1+n_2})\, \mathrm{Im}\left(\frac{3+n_1+in_2}{(1+n_1+in_2)^3}\right).
\end{eqnarray*}

\end{appendix}

\end{document}